\documentclass[aps,prb,twocolumn,showpacs,superscriptaddress,longbibliography]{revtex4-2}  
\usepackage{graphicx}  
\usepackage{dcolumn}   
\usepackage{bm}        
\usepackage{amsmath}
\usepackage{amsthm}
\usepackage{amssymb}
\usepackage{bbold} 
\usepackage{mathrsfs}
\usepackage{array}
\usepackage[normalem]{ulem} 
\usepackage[dvipsnames,usenames]{color}
\usepackage{soul} 
\usepackage{physics}
\usepackage{soul}

\usepackage{color}
\usepackage[dvipsnames,usenames]{color}

\usepackage{braket}
\usepackage{verbatim}
\usepackage{float}
\usepackage{mhchem}
\usepackage[makeroom]{cancel}
\usepackage{upgreek}
\usepackage{commath}
\usepackage{siunitx}
\usepackage{orcidlink}
\usepackage{lipsum}

\usepackage[titletoc]{appendix}

\usepackage{hyperref}
\hypersetup{colorlinks=true, linkcolor=blue, citecolor=blue, urlcolor=blue}

\begin{document}
\title {Inverse determination of light-matter coupling in disordered systems from transmittance spectra}

\author{Thales F. Macedo \orcidlink{0000-0002-8816-2201}}
\affiliation{Instituto de F\'isica, Universidade Federal do Rio de Janeiro, 21941-972 Rio de Janeiro -- RJ, Brazil}
\author{Julián Faúndez \orcidlink{0000-0002-6909-0417}}
\affiliation{Instituto de F\'isica, Universidade Federal do Rio de Janeiro, 21941-972 Rio de Janeiro -- RJ, Brazil}
\affiliation{Departamento de Física y Astronomía, Universidad Andres Bello, Santiago 837-0136, Chile}
\author{Ant\^onio S. Coelho \orcidlink{0000-0002-1860-6743}}
\affiliation{Department of Mechanical Engineering, Federal University of Piauí, 64049-55 Teresina, Piauí -PI, Brazil}
\author{Caio Lewenkopf~\orcidlink{0000-0002-2053-2798}}
\affiliation{Instituto de F\'isica, Universidade Federal do Rio de Janeiro, 21941-972 Rio de Janeiro -- RJ, Brazil}
\author{Mauro S.~Ferreira \orcidlink{0000-0002-0856-9811}}
\affiliation{School of Physics, Trinity College Dublin, Dublin 2, Ireland}
\affiliation{Centre for Research on Adaptive Nanostructures and Nanodevices (CRANN) \& Advanced Materials and Bioengineering Research (AMBER) Centre, Trinity College Dublin, Dublin 2, Ireland}
\author{Felipe A. Pinheiro \orcidlink{0000-0001-8712-0555}}
\affiliation{Instituto de F\'isica, Universidade Federal do Rio de Janeiro, 21941-972 Rio de Janeiro -- RJ, Brazil}
\author{Natanael C. Costa \orcidlink{0000-0003-4285-4672}}
\affiliation{Instituto de F\'isica, Universidade Federal do Rio de Janeiro, 21941-972 Rio de Janeiro -- RJ, Brazil}
\affiliation{Institut f\"ur Theoretische Physik und Astrophysik, Universit\"at W\"urzburg, 97074 W\"urzburg, Germany}


\begin{abstract}
We investigate quantum inverse problems in one-dimensional (1D) electronic disordered systems strongly coupled to optical cavities.
More specifically, we consider the Anderson and the Aubry-Andre-Harper models connected to electronic reservoirs and embedded in a single-mode optical cavity. 
The light-matter interaction enables photon-assisted hopping processes that significantly modify the transmittance spectrum. 
Within the nonequilibrium Green's function formalism, we implement an inversion-based approach capable of accurately extracting the electron-photon coupling strength directly from transmittance spectra. While cavity coupling acts as a minor perturbation within the Anderson model, yielding broad yet precise parameter estimates, its influence is markedly different in the Aubry-Andr\'e-Harper model.
The latter exhibits a sharp metal-insulator transition in 1D, thus resulting in more pronounced cavity-induced spectral changes. 
This renders even more accurate inverse solutions, offering unparalleled precision in the characterization of low-dimensional disordered systems. Altogether, our results demonstrate that the quantum inverse problem provides a robust diagnostic tool for quantum materials, particularly effective for systems exhibiting metal-insulator transitions.
\end{abstract}
\maketitle

\section{Introduction}
\label{sec:intro}

Recent advances in cavity quantum electrodynamics (QED) have opened new avenues for modifying the transport properties of low-dimensional systems through strong light-matter coupling\,\cite{Ritsch2013,Schlawin2019,Schlawin2022,Garcia2021,Lu2025}. Embedding an electronic system in an optical cavity can give rise to photon-assisted processes, renormalization of electronic parameters, and novel hybrid excitations\,\cite{Orgiu2015,Arwas2023}.
Indeed, experiments have demonstrated striking manifestations of these effects, such as supersolid and superradiant Mott insulating phases in ultracold gases \cite{Leonard2017a,Leonard2017b,Landig2016,Klinder2015,Vaidya2018}, light-induced superconductivity in \ce{K3C60} \cite{Budden2021}, cavity-enhanced superconducting correlations in theoretical proposals \cite{Sentef2018}, and enhanced conductivity in disordered organic semiconductors \cite{Orgiu2015}.
Even the vacuum QED fluctuations of a dark cavity can change electronic properties, for instance, 
field fluctuations breakdown the topological protection of the integer quantum Hall effect\, due to induced long-range hoppings \cite{Appugliese2022,Arwas2023,Enkner2025}.
These results highlight the potential of cavity QED to engineer transport phenomena in low-dimensional systems.

Despite this recent progress, the interface between condensed-matter physics and cavity QED remains in its infancy. Even in single-particle problems, the dominant effects of light-matter coupling on material properties -- e.g., for the electronic transport -- are not yet fully established. When disorder is present, the analysis becomes more challenging and the resulting behavior can be difficult to predict. Recent studies indicate that coupling to cavity modes can modify localization lengths in disordered media and, in some cases, open additional transport channels \cite{Hagenmuller2017,Moreno2022,Hagenmuller2018,Arwas2023}. 
However, determining the parameters that control these hybrid systems from experimentally accessible observables -- including the electron-photon coupling strength and, consequently, the effective cavity finesse -- remains an open problem.

The finesse of an optical cavity measures the ratio between the free spectral range and the resonance linewidth using transmission or reflection spectroscopy\,\cite{lu2023}. Environmental perturbations such as thermal drifts, acoustic vibrations, or mechanical instabilities naturally broaden the resonance peaks. Estimating the finesse in cavity QED experiments becomes even more challenging when matter degrees of freedom are strongly coupled to the cavity field. 
When matter is embedded in a cavity within a strong coupling regime, the resulting hybrid quasiparticle resonances reflect nonperturbative renormalizations of the corresponding uncoupled subsystems \cite{menghrajani2024,Fischbach2025,Rider2024,Lalanne2018,Hertzog2019}. In this regime, the measured cavity spectrum no longer reflects the bare finesse, but instead a hybrid linewidth that includes contributions from both photon leakage and light-matter interactions. Similarly, estimating the strength of the light-matter coupling is usually facilitated in the weak and strong coupling regimes, by determining the emitter's lifetime inside the cavity (Purcell effect) and the vacuum Rabi splitting, respectively\,\cite{Kockum2019,Forn2019}. However, in the ultrastrong coupling regime, and for large systems (beyond a single quantum dot), it is not clear how to determine it. These challenges motivate the search for indirect strategies, in which one is able to extract the effective cavity parameters by measuring the properties of the embedded material, such as its conductance.

Bridging this gap requires estimating Hamiltonian parameters directly from experimental data -- that is, solving a quantum inverse problem (QIP). Although controlled implementations in coupled cavity arrays have demonstrated feasibility for QIP\,\cite{Baum2022,Saxena2023,Patton2024}, extending the approach to quantum materials is considerably more difficult due to device variability, uncontrolled disorder, and additional low-energy degrees of freedom, such as phonons.
Within this context, Mukim \textit{et al.}\,\cite{Mukim2020} recently proposed a QIP protocol that infers model parameters directly from energy-resolved conductance fluctuations, proving especially effective for disordered systems. 
This concept was later extended to multi-terminal systems, enabling spatial mapping of local disorder through systematic inversion of the corresponding conductance matrices\,\cite{Mukim2022}, and have been used to explore realistic condensed matter systems, as MoS$_{2}$, AuCl$_{3}$\,\cite{Duarte_2024}, and graphene nanoribbons~\cite{Mukim_2025}. 
Indeed, these works demonstrate that transport measurements may be used as a diagnostic tool capable of resolving disorder strength, and spatial distribution of impurities. 
Thus, applying this methodology to cavity-embedded disordered systems 
has the potential to be a powerful tool to estimate cavity parameters, particularly the light-matter coupling strength, offering a new route toward quantitative characterization of cavity quantum materials.


In this work, we address this issue by systematically employing the QIP method\,\cite{Mukim2020} on two paradigmatic models of wave transport in 1D disordered systems: the Anderson model\,\cite{Anderson1958,Abrahams1979,Lee1985} and the Aubry-Andre-Harper (AAH) model \cite{Harper55,Aubry1980}. 
For both cases, we couple their fermionic degrees of freedom with a single quantized cavity mode, and employ the nonequilibrium Green's functions (NEGF) formalism \cite{Meir1992, Haug2007, Datta1997, Moreno2022} to compute the transmittance response. 
We make use of the aforementioned inversion strategy \cite{Mukim2020, Mukim2022} to recover the input parameters from simulated data. 
This allows us to assess the robustness and precision of the method across regimes ranging from fully localized to fully extended, including the critical states in the AAH model. 

The paper is organized as follows. 
In Section \ref{Sec:model}, we introduce the Anderson model and the AAH model, formulate their interaction with the cavity QED field, and outline the key physical observables under investigation. 
Section \ref{Sec:results} is devoted to a detailed presentation and discussion of the results. The impact and feasibility of our results are discussed in Section \ref{Sec:discussion}, while our concluding remarks are summarized in Section \ref{conclusion}.

\color{black}

\section{Model and methodology}
\label{Sec:model}

\subsection{The Anderson model}

To describe fermions in a 1D disordered chain, we first consider the Anderson model~\cite{Anderson1958,Markos2006}, whose Hamiltonian reads
\begin{equation}
\mathcal{H}_{AM} = -t\sum_{j }(c^{\dagger}_{j}c_{j + 1}^{\phantom{\dagger}}+ \text{H.c})  +  \sum_{j} \epsilon_{j}c^{\dagger}_{j}c_{j}^{\phantom{\dagger}},
\label{eq:and_ham}
\end{equation}
where the sums run over a chain with $L$ sites and $c^{\dagger}_j$ ($c_j$) denotes the creation (annihilation) operator of a spinless fermion at a given site $j$. 
The first term on right-hand side of Eq.\,\eqref{eq:and_ham} describes the kinetic operator, 
where $t$ is the hopping integral between nearest neighbor sites and H.c. stands for Hermitian conjugate. 
The second term corresponds to the disordered on-site potential, with $\epsilon_{\mathbf{i}}$ being a site-dependent random variable uniformly distributed in the interval $[-W/2,W/2]$, where $W$ is the disorder strength.
Hereafter, we set the hopping integral $t\equiv1$ as the model energy scale.

We recall that the 1D Anderson model, even for an infinitesimal amount of disorder $W$, exhibits exponentially localized eigenstates in the thermodynamic limit, a phenomenon known as Anderson localization\,\cite{Markos2006}. 
That is, for sufficiently large system sizes, there is no diffusive transport and all eigenstates are  exponentially localized.
Although no metal–insulator transition occurs in 1D, the Anderson model remains a powerful tool to understand the fundamental role of disorder in quantum systems, especially when novel quantum effects, such as those induced by cavity QED coupling, are incorporated.
This model provides a baseline to assess how cavity coupling affects localization and electronic transport\,\footnote{We note that, beyond electronic realizations, the Anderson Hamiltonian has also been proposed for analog quantum simulation with coupled nanoelectromechanical resonator arrays\,\cite{zalalutdinov2006,lozada2016}.}.

\subsection{The Aubry-Andre-Harper model}\label{AAH_section}

As a second approach to describe disorder in a 1D system, we examine the Aubry-Andre-Harper (AAH) model~\cite{Aubry1980,Harper55}, whose Hamiltonian reads
\begin{align}
\nonumber
\mathcal{H}_{AAH} = &-t\sum_{j }(c^{\dagger}_{j}c_{j + 1}^{\phantom{\dagger}}+ \text{H.c}) \\
& +  V\sum_{j}\cos{( 2\pi \beta j + \phi)}c^{\dagger}_{j}c_{j}^{\phantom{\dagger}}.
\label{eq:AAH-Hamiltonian}
\end{align}
As before, the first term denotes the kinetic operator, while the second describes an on-site quasiperiodic potential of amplitude $V$. 
We set $\beta = (\sqrt{5}-1)/2$ (the inverse golden ratio), so that the modulation is incommensurate with the lattice; $\phi$ is an arbitrary global phase.

In contrast to the 1D Anderson model, where any finite uncorrelated disorder localizes all single-particle states, the quasiperiodic potential in the AAH model leads to a self-dual property. In other words, under a discrete Fourier transform, the Hamiltonian maps onto itself while exchanging the hopping amplitude and the potential strength\,\cite{Dominguez2019}. In view of this, the spectrum exhibits a localization transition at the self-dual point $V=2t$ for all single-particle eigenstates; they are extended for $V<2t$ and localized for $V>2t$, while at $V=2t$ they are critical and display multifractal features\,\cite{Evangelou2000,Dominguez2019,macedo2024}. 
Because it yields a metal-insulator transition in 1D, the AAH model serves as an important platform for investigating light–matter coupling in QED cavities beyond the single-band Anderson model. Furthermore, its sharply defined localization transition makes it fertile ground for exploring cavity-induced effects near critical points.

\subsection{Cavity effects}

\begin{figure}[t]
\includegraphics[width=0.9\linewidth]{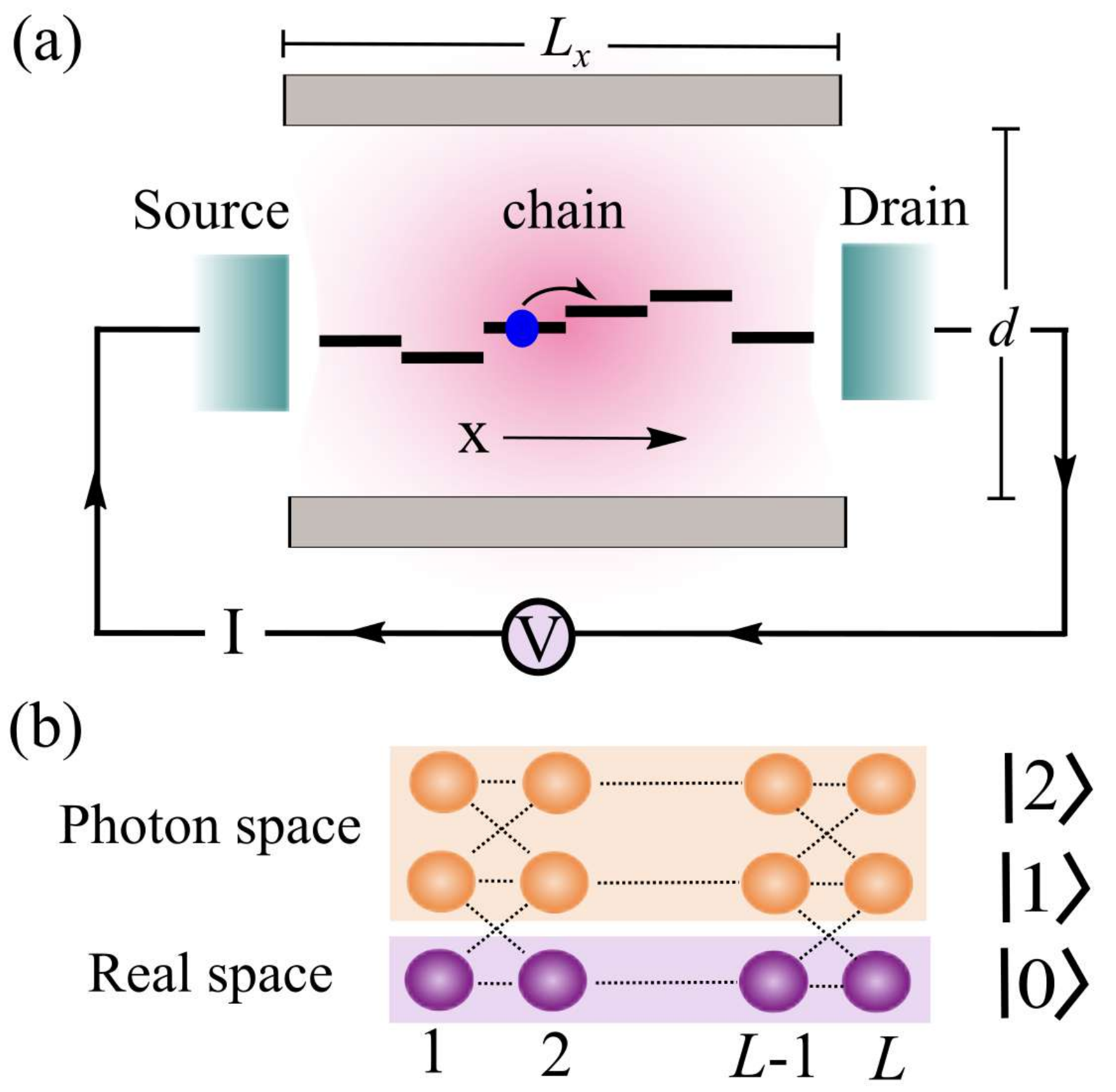}
\caption{(a) Schematic illustration of a 1D chain embedded in an optical cavity. Each horizontal line corresponds to a lattice site, with its vertical position indicating the photon number ($N$) subspace. Here, $\gamma$ defines the light-matter interaction. The 1D chain is terminated at both ends by semi-infinite leads coupled to thermal reservoirs in thermal equilibrium, which serve as source and drain, respectively.
(b) Electron hopping processes are accompanied by photon emission or absorption, inducing transitions between photon number subspaces $\ket{N}$ and giving rise to an effective multi-chain structure in the combined electron-photon Hilbert space. }
\label{fig:placeholder}
\end{figure}

The interaction between the electromagnetic (EM) field and the electronic degrees of freedom is given by the minimal coupling, $ \mathbf{p}\to\mathbf{p}-q\mathbf{A}(\mathbf{r},t)$, with $\mathbf{A}$ being the vector potential and $q$ the particles' charge.
For lattice models, this is cast by the Peierls substitution\,\cite{Peierls1933,marder2010condensed, Cresti2021},
$$ t_{\mathbf{i}\mathbf{j}}\to t_{\mathbf{i}\mathbf{j}}\exp\!\Big[\frac{{\rm i} q}{\hbar}\int_{\mathbf{r}_i}^{\mathbf{r}_j}\mathbf{A}\!\cdot d\boldsymbol{\ell}\Big],$$
which preserves gauge invariance \cite{Jiajun2020}.
Here we work in the long-wavelength (dipole approximation) limit, in which 
the EM wavelength is much larger than the lattice spacing, $\lambda_{\mathrm{EM}}\gg a$, so that $\mathbf{A}(\mathbf{r},t)$ is approximately uniform. 
Therefore, for 1D systems with nearest neighbor hoppings, the Peierls substitution reduces to $t\rightarrow t\,\exp\!\left[{\rm i}\frac{q a}{\hbar}A_x\right]$.

We now couple the 1D disordered chain to a QED cavity.
The cavity can be modeled as two parallel mirrors of linear size $L_x \to \infty$ (or $L_x \gg a$), separated by a finite distance $d$; see, for instance, Fig.\,\ref{fig:placeholder}\,(a). This geometry leads to a continuum of EM modes along the $q_x$ direction, while the modes along $q_z$ are quantized. As discussed in Ref.\,\onlinecite{Li2022}, one may introduce an ultraviolet cut-off for the EM modes, and deal with a macroscopically effective light-matter coupling term. Thus, to simplify our analysis, we examine the effects of a single, linearly polarized, effective mode, treating it as quantized, i.e.~$A_x = A_0 \left(b^{\dagger} + b\right)$,
with $b^{\dagger}$ and $b$ being the photon creation and annihilation operators, respectively. $A_0$ sets the light-matter coupling strength, which depends on the mode volume and its frequency.

The Hamiltonian of the Anderson model coupled to a QED-cavity reads
\begin{align}
    \mathcal{H}_{AM,c} = & - t \sum_{ j } \bigg[ e^{\mathrm{i}\gamma(b + b^{\dagger}) } c^{\dagger}_{j} c_{j+1} + {\rm H.c.}\bigg] \\
    &+ \sum_{j} \epsilon_{j} c^{\dagger}_{j}c_{j} + \hbar \omega_{0} b^{\dagger} b
 \label{eq:cav_ham}
\end{align}
and, similarly, the Hamiltonian of the AAH model becomes 
\begin{align}
\nonumber
\mathcal{H}_{AAH,c} = & -t \sum_{ j } \bigg[ e^{\mathrm{i}\gamma(b + b^{\dagger}) } c^{\dagger}_{j} c_{j+1} + {\rm H.c.}\bigg]  \\
& +  V\sum_{j}\cos{( 2\pi \beta j + \phi)}c^{\dagger}_{j}c_{j} + \hbar \omega_{0} b^{\dagger} b , 
    \label{AAHcavity}
\end{align}
where $\gamma=qaA_{0}/\hbar$ quantifies the light-matter coupling strength, and the last term denotes the bare photon energy, $\hbar \omega_{0}$.
Notice that the kinetic terms of these Hamiltonians have an operator ``$e^{\mathrm{i}\gamma(b + b^{\dagger}) }$'' which couples electronic hopping to photon emission and absorption, as illustrated in Fig.\,\ref{fig:placeholder}\,(a). This dressing modifies the nearest-neighbor hopping amplitude and can generate effective long-range terms through virtual photon processes\,\cite{Mikami2016,Jiajun2020,Li2022}. 
We also stress that, since we are dealing with a single-electron problem, a $1/\sqrt{L}$ factor in the electron-photon coupling is not explicitly required to enforce the extensivity of the energy, being already incorporated in the $A_0$ constant.

The total Hilbert space is spanned by the tensor product states $\ket{j, N}$, where $j = \{1, \dots, L\}$ denotes the chain site and $N = \{0, \dots, N_{\text{ph}}\}$ is the photon number.
Within this basis, the Baker-Campbell-Hausdorff formula allows us to obtain the Hamiltonian matrix elements between photon number sectors, $\langle M|\mathcal{H}|N\rangle$. More details are discussed in Refs.\,\onlinecite{Moreno2022,macedo2024}.
Therefore, the problem may be viewed as a set of coupled chains indexed by $N$, as shown in Fig.\,\ref{fig:placeholder}\,(b), where the links between chains with different photon labels (i.e., the off-diagonal matrix elements between different photon-number sectors) lead to photon-assisted tunneling processes.
This formulation allows for an exact treatment of the light-matter interaction up to a cut-off in the total number of photons, $N_{\text{ph}}$.

\subsection{Transport properties}

To investigate the electronic transport properties, the disordered chain is treated as a finite system coupled to two semi-infinite leads acting as equilibrium reservoirs, as illustrated in Fig.\,\ref{fig:placeholder}\,(a).
The influence of the leads on the central region is cast by embedding self-energy corrections, resulting in an effective non-Hermitian Hamiltonian.
These self-energies, denoted by $\Sigma_\text{S}(E)$ and $\Sigma_\text{D}(E)$, describe the exchange of particles between the 1D chain and the source and drain leads, respectively \cite{Datta2005}.
In the absence of photons, and for an 1D chain of linear size $L$ in the central region, the expressions for the leads' self-energies are exact, with matrix elements given by
\begin{align}
\left[ \Sigma_{\text{S}}^{\rm r} (E)\right]_{j,j'} & = \delta_{j,1} \delta_{j',1} \frac{t^{2}_{\rm c}}{t_{\rm S}} f\left( \frac{E - \mu_{\rm S}}{t_{\rm S}} \right) \\
\left[ \Sigma_{\text{D}}^{\rm r}(E) \right]_{j,j'} & = \delta_{j,L} \delta_{j',L} \frac{t^{2}_{\rm c}}{t_{\rm D}} f\left( \frac{E - \mu_{\rm D}}{t_{\rm D}} \right)~,
\end{align}
where $$f(x) = \frac{2}{x + {\rm i}\sqrt{4 - x^2}}~,$$ and the superscript $r$ denotes the retarded.
Here, $t_{\rm S/D}$ and $\mu_{\rm S/D}$ correspond to the hopping integrals and the Fermi levels in the source (S) and drain (D),
respectively, while $t_{\rm c}$ is the hopping between the leads and the central chain; $E$ sets the energy of the electronic modes in the leads. 
We take symmetric lead parameters, $t_{S}=t_{D}=t_{c}=t$ for simplicity, to reduce the parameter space and isolate the impact of the light–matter coupling. 
In addition, we set $\mu_{S}=\mu_{D}=0$ to address the linear response regime typical of quantum transport experiments.
Introducing more general lead parameters would increase the number of independent variables and may influence the quantitative performance of the inversion procedure by modifying the transmittance profile. 
However, from a qualitative viewpoint, it does not compromise the ability of the method to recover the Hamiltonian parameters.

Following Ref.\,\onlinecite{Moreno2022}, we extend the definitions to the cavity setup by introducing the lead self-energy $\tilde{\Sigma}_{\alpha}^{\mathrm{r}}(E)$ in the product basis ${|j,N \rangle}$. Its matrix elements are
\begin{equation}
\langle j,N \vert \tilde{\Sigma}_{\alpha}^{\mathrm{r}}(E) \vert j',M \rangle
= \delta_{N,M} ~ \bigl[\Sigma_{\alpha}^{\mathrm{r}}(E - N\hbar\omega_{0})\bigr]_{j j'}~,
\end{equation}
with $\alpha = {S,D}$ and $\omega_{0}$ being the cavity mode frequency. It corresponds to the case of an electron with energy $\varepsilon \equiv E - N\hbar\omega_{0}$ in the leads when the cavity hosts $N$ photons. Here, we assume that electron-photon coupling is absent in the leads (i.e., it acts only in the central region), and that the tunneling between leads and chain is independent of the number of photons within the cavity.
Given this, the dressed retarded and advanced Green’s functions are
\begin{equation}
    G^{r,a}(E) = \bigl[\,E \mathbb{1} - \mathcal{H} - \tilde{\Sigma}^{r,a}(E)\,\bigr]^{-1},
\end{equation}
where $\tilde{\Sigma}^{r,a}(E) = \tilde{\Sigma}^{r,a}_\text{S}(E) + \tilde{\Sigma}^{r,a}_\text{D}(E)$, $E$ is the electron energy, and $\mathbb{1}$ is the identity operator.

In the linear-response regime, the conductance is directly computed from the energy-resolved transmittance obtained from the nonequilibrium Green's functions formalism,
\begin{align} 
\label{eq:landauerbutiker}
\nonumber   \mathcal{T}_{NM}(E) = \Trace \big[ \bra{N} \Gamma_\text{S} \ket{N} \bra{N} G^{r}(E) \ket{M} \times \\ \bra{M} \Gamma_\text{D} \ket{M} \bra{M} G^{a}(E) \ket{N} \big],
\end{align}
where the level-broadening matrices are defined as
\begin{equation}
  \Gamma_{\alpha}(E)
  = {\rm i} \left[ \tilde{\Sigma}_{\alpha}^{r}(E)
  - \tilde{\Sigma}_{\alpha}^{a}(E) \right]~,
\end{equation}
with $\alpha = \text{S},\text{D}$.
Here, $\mathcal{T}_{NM}(E)$ denotes the transmission probability for a process in which an electron enters from the source with energy $\varepsilon = E - N\hbar\omega_{0}$ while the cavity hosts $N$ photons, and exits into the drain with energy $\varepsilon' = E - M\hbar\omega_{0}$ with the cavity having $M$ photons. Thus the total energy is conserved, $\varepsilon + N\hbar\omega_{0} = \varepsilon' + M\hbar\omega_{0} = E$. We emphasize that virtual photon emission and absorption processes are incorporated through the dressed Green's functions, so $\mathcal{T}_{NM}(E)$ includes both real and virtual photon-assisted contributions.
Since at $T=0$ no real cavity photons are excited, then $\mathcal{T}_{00}(E)$ provides the leading contribution to the transmission.
Therefore, in what follows, we restrict our analysis to $N=M=0$, so that $\mathcal{T}(E) \equiv \mathcal{T}_{00}(E)$. In this regime, the conductance reads $G(\mu) = (e^2/h)\mathcal{T}(\mu)$, the electronic transport can be viewed as an elastic scattering process, and the NEGF conductance is described by the Landauer-Büttiker formula \cite{Hernandez2007}.
We stress that this framework can be readily formulated for multi-terminal systems with arbitrary geometries \cite{Mukim2022}. 
For isolated systems subject to open or periodic boundary conditions, where transmittance is not well-defined, the inversion scheme can in principle still be applied by defining the cost function in terms of alternative observables, such as the density of states or the spectral function. 
These quantities similarly reflect the photon-assisted hybridization and disorder-induced effects that characterize the system’s electronic structure.

\subsection{The quantum inverse problem approach}
\label{subsec:IP}

Characterizing disordered systems is challenging because typically physical observables depend sensitively on the microscopic disorder configurations. 
The ``direct'' problem (i.e., computing the conductance for a known Hamiltonian) is, in principle, straightforward. However, inferring model parameters from conductance data is typically ill posed. 
Promising developments to perform QIP from conductance data are discussed in Refs.~\onlinecite{Mukim2020,Mukim2022}, from which we adapt the approach we use to analyze the cavity-coupled systems presented in Sec.\,\ref{Sec:model}. 
For completeness, we discuss the methodology here.

The QIP method introduced in Refs.\,\onlinecite{Mukim2020,Mukim2022} relies on a misfit function, $\chi(\boldsymbol{\Omega})$, which quantifies the difference between the energy-dependent transmittance spectra (or its logarithm) of a given experimental realization, $\mathcal{T}_{\mathrm{true}}(E)$, and the configurationally averaged model prediction $\big\langle \mathcal{T}(E;\boldsymbol{\Omega}) \big\rangle$ evaluated for a candidate parameter set $\boldsymbol{\Omega}$ (e.g., disorder strength and cavity coupling).
The misfit function is defined over an energy window $[\mathcal{E}_-, \mathcal{E}_+]$ as
\begin{equation}
    \chi(\boldsymbol{\Omega}) = \frac{1}{(\mathcal{E}_{+} - \mathcal{E}_{-})} \int_{\mathcal{E}_-}^{\mathcal{E}_+} \dd{E} \Big[\mathcal{T}_{\mathrm{true}}(E) - \big\langle \mathcal{T}(E;\boldsymbol{\Omega}) \big\rangle \Big]^2.
\label{eq:misfit_func}
\end{equation}
By minimizing $\chi(\boldsymbol{\Omega})$ over the admissible domain, thereby achieving maximal agreement with the observed conductance data, one obtains the best parameter estimate $\boldsymbol{\Omega}_{\rm min}$.
Here $\big\langle \mathcal{T}(E;\boldsymbol{\Omega}) \big\rangle$ is the average of the model transmittance over $N_{\rm dis}$ disorder  configurations, typically  $N_{\rm dis} \sim 10^3$.
Being a loss function, $\chi(\Omega)$ quantifies the deviation between the trial input function and its configurationally averaged counterpart. If this deviation were evaluated at a single energy, many distinct solutions would satisfy the criterion, providing little meaningful information about the underlying structure. By contrast, assessing the misfit over a broader energy range allows us to exploit spectral features that are not apparent at a single energy value. This additional information effectively constrains the inversion and enables the correct solution to be uniquely identified.

For practical implementations, we evaluate the averaged misfit function over $N_{r}$ independent realizations. 
That is, for each $\mathcal{T}^r_{\mathrm{true}}$ realization ($r=1,\dots,N_{r}$), we compute $\chi_{r}(\boldsymbol{\Omega})$ defined in Eq.\,\eqref{eq:misfit_func}, and then obtain its mean value
$$\overline{\chi}(\boldsymbol{\Omega})=\frac{1}{N_{r}}\sum_{r=1}^{N_r}\chi_{r}(\boldsymbol{\Omega})$$
and standard deviation error bars.
In addition, because localized states yield strongly suppressed transmittance, we evaluate the misfit function for $\ln(\mathcal{T}(E))$ rather than $\mathcal{T}(E)$. This choice enhances the misfit response for such states, while preserving its definition. Finally, in the case of the cavity QED, the QIP is formulated with $\boldsymbol{\Omega} =\{\gamma, W \}$ as the unknown quantity to be extracted from the inversion procedure.

\subsection{The cut-off in the number of photons $N_{ \text{ph} }$}

At this point, it is important to examine the photon-number cut-off $N_{\text{ph}}$, which should be chosen so that it does not affect the accuracy of the misfit function. We recall that $\mathcal{T}_{00}(E)$ channel retains only the low-energy states of the full Hamiltonians in Eqs.\,\eqref{eq:cav_ham} and \eqref{AAHcavity}. Because these states are weakly corrected from sectors with large photon number, we expect that a small cut-off for $N_{\text{ph}}$ is required for accurate results, as discussed below.

To quantify photon-number corrections and determine the cutoff $N_{\text{ph}}$, we analyze the behavior of $\langle \ln[\mathcal{T}(E;\boldsymbol{\Omega})] \rangle$ as a function of $N_{\text{ph}}$. We compare each value to a reference computed with a sufficiently large photon number, used as an approximation to the asymptotic $N_{\text{ph}} \to \infty$ limit. Specifically, we evaluate the relative error of the energy-averaged transmittance over a fixed window,
\begin{equation}
\delta (N_{\text{ph}}) = \frac{\delta I(N_{\rm ph})}{|I_{\text{ref}}|}~,
\label{Eq:delta}
\end{equation}
with
\begin{equation}
I_{\rm ref} =  \int_{\mathcal{E}_-}^{\mathcal{E}_+} \dd{E}\, \big\langle \ln[\mathcal{T}(E;\boldsymbol{\Omega})] \big\rangle_{\rm ref},
\label{eq:Int_error1}
\end{equation}
and $\big\langle \ln[\mathcal{T}(E;\boldsymbol{\Omega})] \big\rangle_{\rm ref}$ taken for fixed $N_{\text{ph}}=10$, while
\begin{align}
\nonumber \delta I(N_{\rm ph}) =   \int_{\mathcal{E}_-}^{\mathcal{E}_+} \dd{E}\, &| \big\langle \ln[\mathcal{T}(E;\boldsymbol{\Omega})] \big\rangle_{\rm ref} \\
& - \big\langle \ln[\mathcal{T}(E;\boldsymbol{\Omega},N_{\rm ph})] \big\rangle |.
\label{eq:Int_error2}
\end{align}
For consistency with the following results, we defined $\mathcal{E}_-=0$ and $\mathcal{E}_-=2t$ in the integrals of Eqs.\,\eqref{eq:Int_error1} and \eqref{eq:Int_error2}.

\begin{figure}[t]
\includegraphics[width=0.4\textwidth]{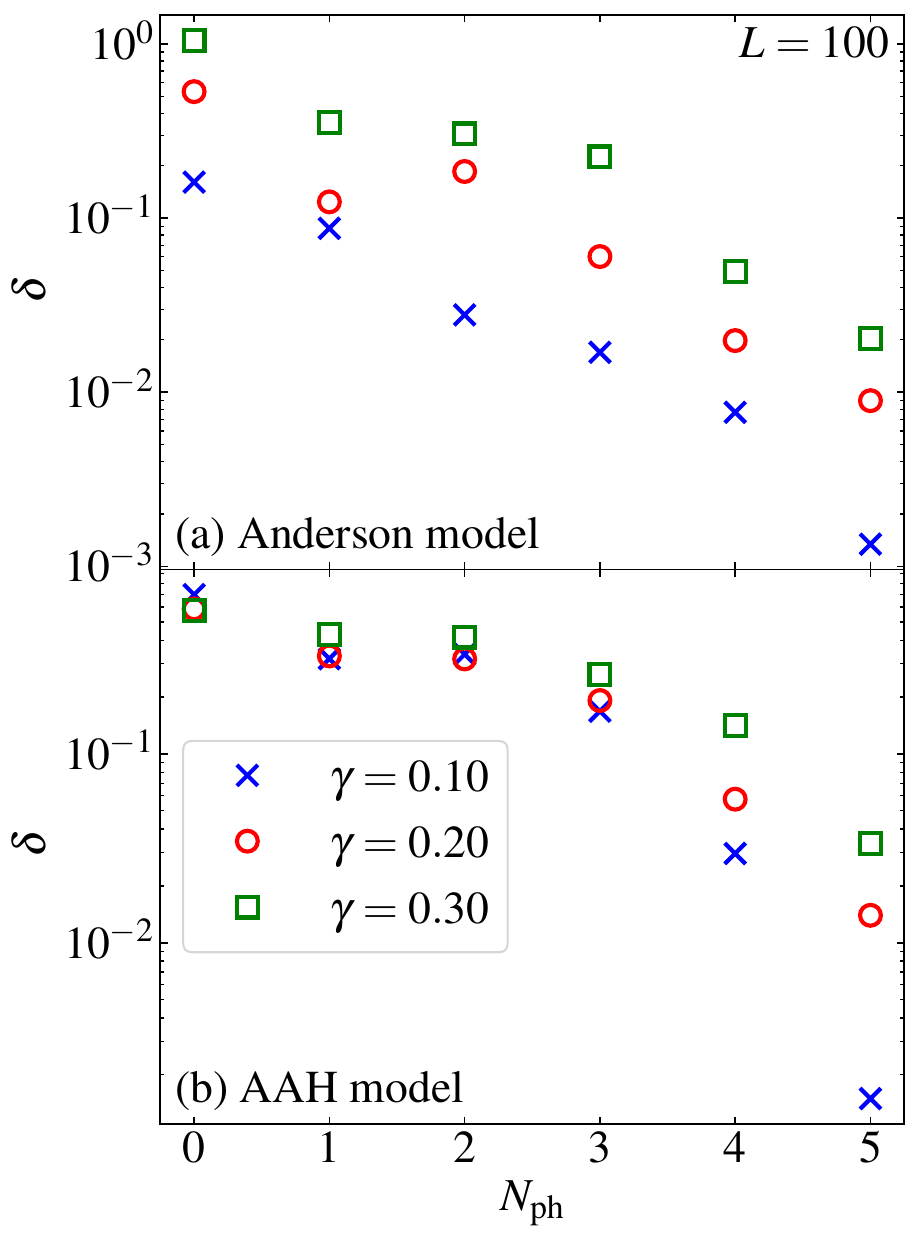}
\caption{Relative error for the integrated transmittance $\big\langle \ln[\mathcal{T}(E;\boldsymbol{\Omega})] \big\rangle$ as a function of the number of photons $N_{\rm ph}$ for (a) the Anderson model, and (b) the AAH model, at fixed $L=100$ sites. 
The results are shown for $\gamma = 0.10$, $0.20$, and $0.30$. 
Here, we set $N_{\rm ph}=10$ as the reference result. }
\label{Fig:delta}
\end{figure}

The behavior of $\delta(N_{\rm ph})$ is displayed in Fig. 2 for (a) the Anderson model and (b) the AAH model. In this figure, the relative error is evaluated using $N_{\rm ph}=10$ as the reference cutoff.
The strong decrease of $\delta(N_{\rm ph})$ with increasing $N_{\rm ph}$ indicates that the integrated transmittance is already well converged in the parameter regime under study. Further increasing $N_{\rm ph}$ primarily enlarges the total Hilbert-space dimension, $D_{\mathrm{total}} = D_{e}(N_{\rm ph}+1)$, and thus the computational cost, without leading to any {qualitative} change in the results. 
Here, $D_{e}$ denotes the electronic contribution to the Hilbert space \cite{Moldoveanu2019} In both models, we analyze $\delta(N_{\rm ph})$ at electron-photon coupling $\gamma = 0.20$ and fixed system size $L=100$. 
We {notice} that the corrections to the integral of $\ln[{\cal T}(E; \boldsymbol{\Omega)}]$ become negligible once the photon cutoff reaches $N_{\rm ph}\geq 5$, where the relative error is about $1\%$ or smaller. 
As $\gamma$ is reduced, the electron-photon interaction effects weaken. 
Correspondingly, $\delta(N_{\rm ph})$ is expected to vanish in the limit $\gamma \to 0$. Consequently, we set the photon cutoof to $N_{\rm ph}=5$, which offers a suitable balance between numerical precision and computational {cost}. 
We also emphasize that other transmittance channels $T_{NM}(E)$, with $N,M \neq 0$, may involve higher-energy states and thus demand a larger photon cutoff, but such channels are typically less relevant for low-temperature transport.

\section{Results}
\label{Sec:results}

\subsection{The Anderson Model}\label{subsec:Anderson}

We begin our analysis of the QIP procedure by considering the Anderson model. According to Ref.~\onlinecite{Moreno2022}, the cavity field mainly renormalizes the localization length $\xi$ of the single-particle eigenstates in the 1D Anderson model. Because this model does not exhibit a metal-insulator transition in 1D (the eigenstates are localized for arbitrarily weak disorder), it provides a convenient testbed for assessing the QIP approach in a regime dominated by localized states.

\begin{figure}
    \centering
    \includegraphics[width=0.45\textwidth]{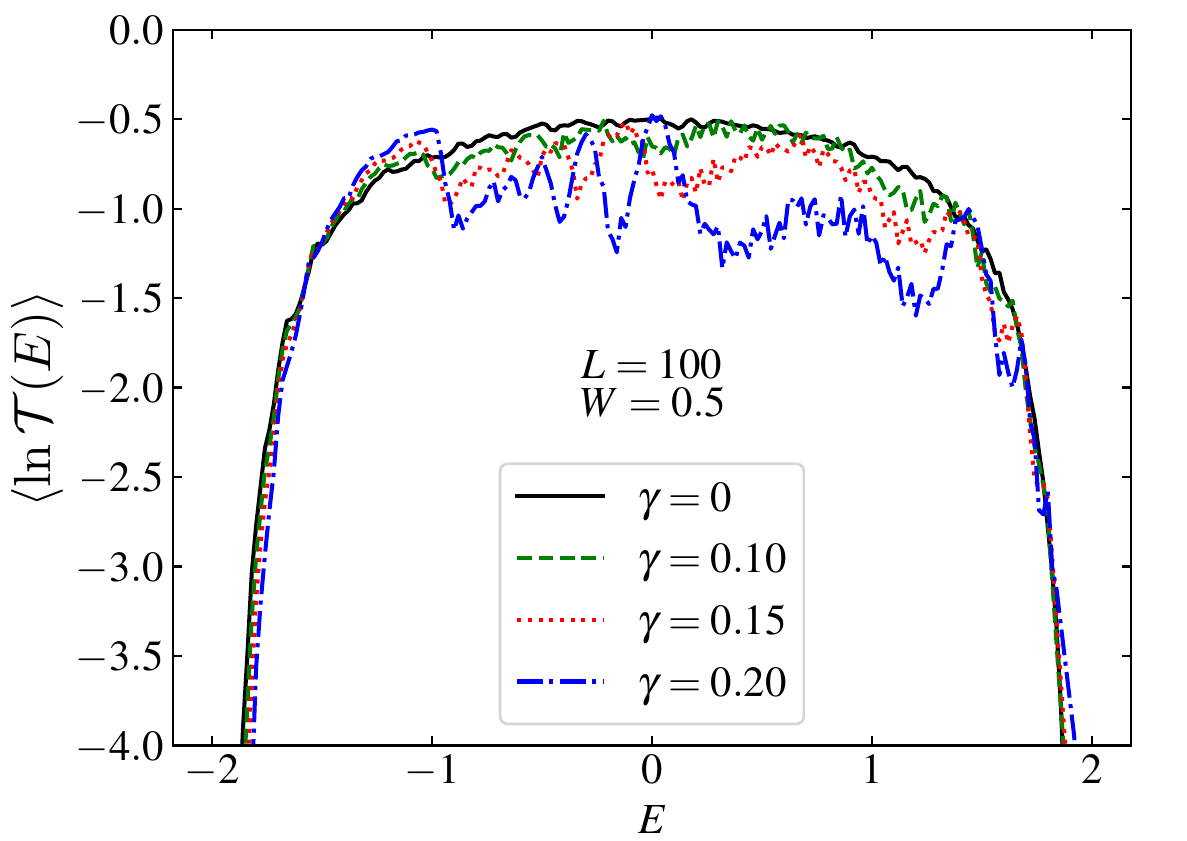}
    \caption{Disorder-averaged logarithm of the transmittance, $\langle \ln \mathcal{T}(E)\rangle$, as function of the energy for $L=100$, disorder strength $W=0.5$, and different electron–photon couplings. In what follows, we define $\gamma_{\rm true}=0.15$ and $W_{\rm true}=0.5$.}
    \label{fig:transmittance}
\end{figure}

As outlined in Sec.\,\ref{subsec:IP}, the QIP approach estimates experimental parameters by inverting the response of the conductance computed within a model, by using a misfit function $\chi$. In the absence of experimental data, we benchmark the procedure using independently blind input data. For the cavity-QED setting, we treat the electron-photon coupling $\gamma$ and the disorder strength $W$ as the unknowns to be inferred. Unless stated otherwise, the synthetic ``experimental/true'' values are $\gamma_{\rm true}=0.15$ and $W_{\rm true}=0.5$.
Obviously, we are not allowed to avail of that information when searching for their values through the inversion process.
In what follows, we also have fixed the photon energy $\hbar \omega_{0}/t = 1$.

We start by fixing the disorder strength to its reference value $W = W_{\rm true}$, so that the misfit function depends only on $\gamma$, i.e.~$ \chi \equiv \chi(\gamma)$.
To highlight the influence of electron-photon coupling on the transport properties, Fig.\,\ref{fig:transmittance} shows the configurationally averaged transmittance $\langle \ln \mathcal{T}(E;\gamma,W_{\rm true}) \rangle$ across the energy spectrum for several values of $\gamma$. 
Each curve is obtained by averaging over $N_{\rm dis}=1000$ disorder realizations.
Two features of Fig.~\ref{fig:transmittance} are particularly noteworthy. 
First, the spectrum of $\langle \ln \mathcal{T} \rangle$ exhibits a high sensitivity to the coupling strength: 
Even a small variation in $\gamma$ leads to significant modifications in the transmittance profile. 
This sensitivity ensures that the misfit function effectively captures changes in the light-matter interaction.
Second, the transmittance shows only minor changes for $E/t \lesssim -1$ when $\gamma$ is varied, which is a subtle point that demands further discussion, since it may affect the effectiveness of the misfit function.

\begin{figure}[t]
\includegraphics[width=0.4\textwidth]{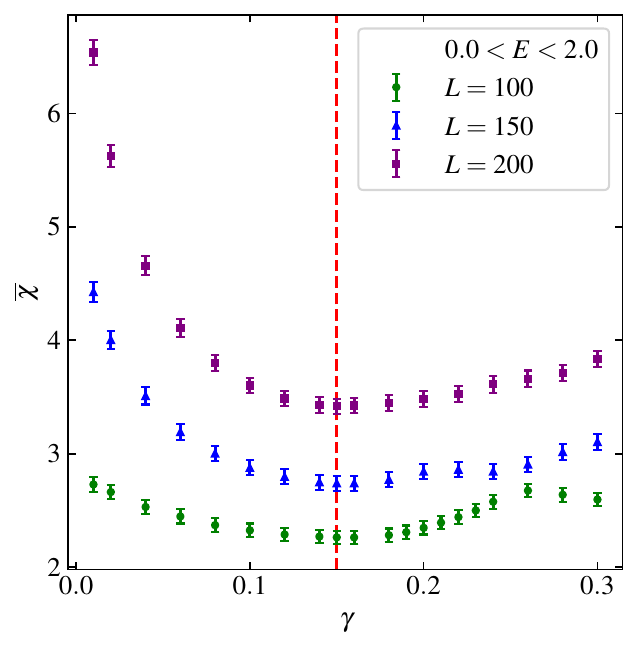}
\caption{Averaged misfit function $\overline{ \chi }$ (in arbitrary units) as a function of the electron-photon coupling $\gamma$ for the 1D Anderson model evaluated in a fixed disorder strength $W = 0.5$. The energy integration in the misfit function is performed over the window  $0 < E < 2t$.  The vertical dashed line highlights the true coupling $\gamma_{\rm true} = 0.15$, where the minimum of $\overline{ \chi }$ occurs.}
\label{fig:chi_vs_gamma}
\end{figure}

For QIP in disordered systems, the choice of the energy-integration bounds $\mathcal{E}_{-}$ and $\mathcal{E}_{+}$ in Eq.\,\eqref{eq:misfit_func} is often arbitrary, since the effects of the disordered potential is normally observed across the entire energy band. In fact, inversion errors were deemed acceptable for windows spanning at least $20\%$ of the bandwidth regardless of their location within the band \cite{Mukim2020}, suggesting that rigid shifts of $\mathcal{E}_{\pm}$ typically produce only minor variations in $\overline{ \chi }$.
However, the situation is different in cavity QED, especially when $\hbar\omega_{0}$ is smaller than the bare electronic bandwidth (which is our case in this work). As shown in Ref.\,\onlinecite{Moreno2022} for the 1D Anderson model, when $\hbar\omega_{0}=2t$ the dominant cavity-induced modifications of $\mathcal{T}(E)$ occur near the top of the band ($E \approx +2t$).
This behavior arises from the hybridization of nearly degenerate electronic states belonging to  different photon-number sectors \cite{Moldoveanu2019}. 
In the enlarged electron-photon Hilbert space, the unperturbed energy levels are given by $E_{\alpha,n} = E_{\alpha} + n\hbar \omega_0$, where $E_{\alpha}$ is the electronic energy and $n$ is the photon occupation. 
The light-matter interaction hybridizes these states more effectively when the resonance condition $E_{\alpha} - E_{\beta} \approx (m-n)\hbar\omega_0$ is satisfied. 
This coupling induces avoided crossings that break the $E \rightarrow -E$ symmetry inherent to the isolated 1D Anderson model. 
In our model, this effect is particularly pronounced near the upper band edge, where the electronic density of states and the photon frequency conspire to satisfy the resonance condition most effectively. 
Conversely, states near the lower band edge lack nearly degenerate counterparts in relevant higher-order photon sectors, resulting in negligible corrections to the transmittance, as shown in Fig.~\ref{fig:transmittance}.
Similar features have also been seen in the case of edge disorder in 2D nanoribbons, where the inversion accuracy appears to be sensitive to the position of the energy window \cite{Mukim_2025}.

\begin{figure}
    \centering
    \includegraphics[width=0.38\textwidth]{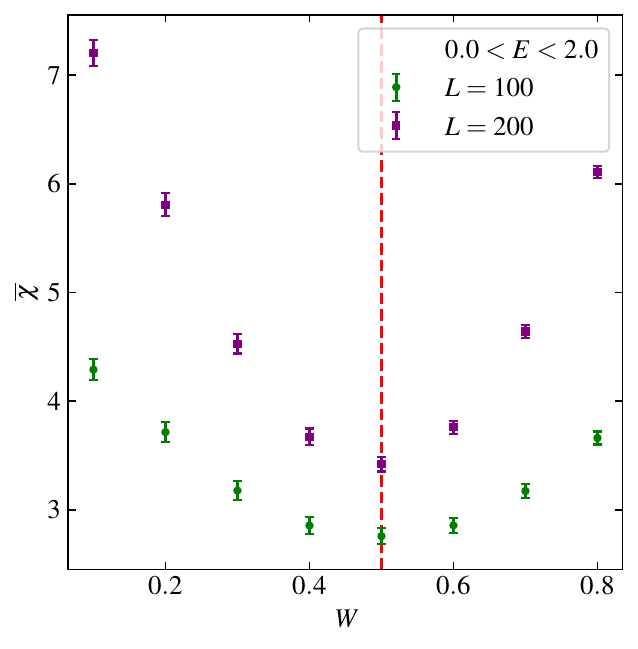}
    \caption{Averaged misfit function $\overline{ \chi }$ (in arbitrary units) as a function of disorder strength $W$, for the 1D Anderson model, computed for a fixed electron-photon coupling $\gamma = 0.15$. The energy integration in the misfit function is performed over the window $0 < E < 2t$. The vertical dashed line highlights the true $W_{\rm true} = 0.5$, where the minimum of $\overline{ \chi }$ occurs.}
    \label{fig:chi_vs_W}
\end{figure}

The above discussion has a practical consequence: if the integration window in Eq.\,\eqref{eq:misfit_func} is chosen near the bottom of the band ($ \approx -2t$), $\overline{ \chi }$ would develop a very shallow minimum, making the QIP less effective to determine the cavity settings (in principle, it would require an exponentially larger number of measurements $N_{r}$ to determine the minimum). This situation is analogous to the case where $\hbar\omega_{0}$ exceeds the electronic bandwidth, which yields only small variations of $\mathcal{T}(E)$ across the band. In both cases, clear effects in the misfit function are expected only in the ultrastrong-coupling regime (i.e., very large $\gamma$), which is challenging experimentally. As discussed below, gate-induced experiments in QED cavities typically involve low photon energies\,\cite{Delbecq2011,Deng2015,Deng2016,Physical2018}. Thus, in line with the results in Fig.\,\ref{fig:transmittance}, it is preferable to choose the energy-integration bounds $\mathcal{E}_{-}$ and $\mathcal{E}_{+}$ close to the top of the band. In view of this, we evaluate the integral in Eq.\,\eqref{eq:misfit_func} with $\mathcal{E}_{-}=0$ and $\mathcal{E}_{+}=2t$. This choice covers the energy range where cavity effects are most pronounced and monitors the sensitivity of $\overline{ \chi }$ to small variations of these bounds.

With the above setup in mind, Fig.\,\ref{fig:chi_vs_gamma} displays the behavior of the averaged misfit function $\overline{ \chi }$ as a function of $\gamma$, for a fixed disorder strength $W = W_{\rm true} = 0.5$, and different system sizes. 
Interestingly, it exhibits a clear minimum near the reference coupling $\gamma=0.15$ (vertical dashed red line). Over the range of $L$ examined, the position of the minimum shows weak dependence on system size. By contrast, the minimum 
becomes more pronounced as the system size increases. 
This trend is consistent with an intrinsic property of the model rather than an artifact of the IP methodology: for larger systems, the exponential localization of the wavefunction is captured more faithfully, which sharpens the minimum. 
Apart from this, the results in Fig.\,\ref{fig:chi_vs_gamma} clearly indicate that the misfit function provides a practical estimator of the electron-photon coupling, $\gamma$. 

\begin{figure}
    \centering
    \includegraphics[width=0.48\textwidth]{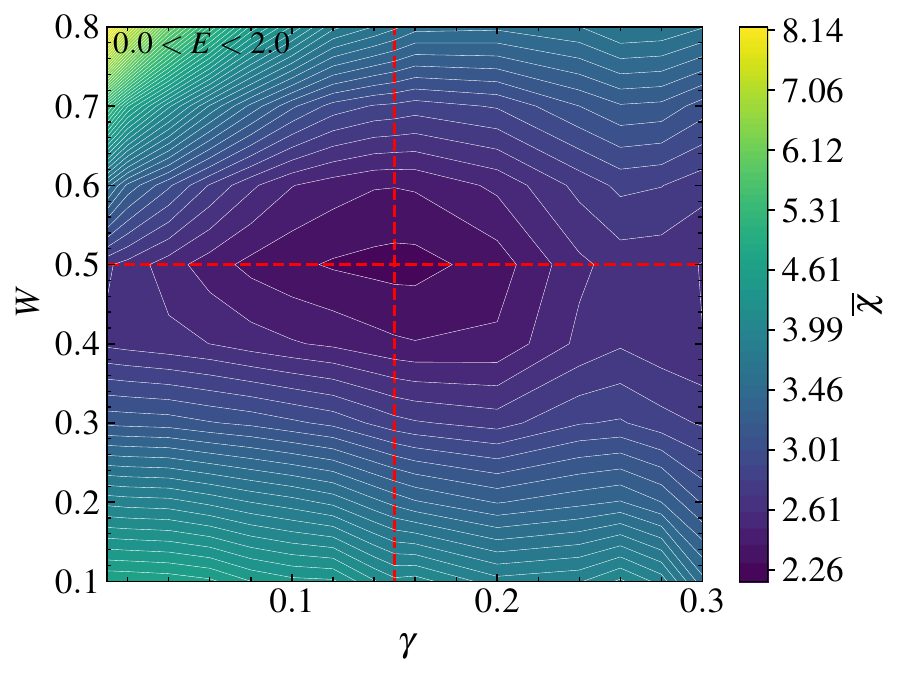}
    \caption{Contour map of the averaged misfit function $\overline{ \chi }(\gamma, W)$ (in arbitrary units) for the 1D Anderson model. The system is fixed at $L = 100$ and $\overline{ \chi }$ is evaluated over the energy window $0 < E < 2$. The color bar indicates the intensity of $\overline{ \chi }$ while the crossing of the dashed lines determines the true parameters $(\gamma_{\rm true}, W_{\rm true}) = (0.15, 0.5)$.}
    \label{fig:contour_plot}
\end{figure}

Having validated the method’s ability to retrieve the coupling constant 
$\gamma$, we next test its sensitivity to assess the disorder strength $W$.
Similarly the previous analysis, we
fix $\gamma=\gamma_{\rm true}=0.15$ and vary $W$.
The result for $\overline{ \chi }(W)$ is presented in Fig.\,\ref{fig:chi_vs_W} for several system sizes. The curves display pronounced minima at $W \approx W_{\rm true}=0.5$, highlighted by the vertical dashed red line, with a weak size dependence. 
Taken together, these results confirm that the QIP protocol enables accurate estimation for both $W$ and $\gamma$ from conductance data.
The shallow minima in $\overline{ \chi }(\gamma)$ at fixed disorder strength, shown in Fig.\,\ref{fig:chi_vs_gamma}, is due to weakly localized states. 
For larger $W$, where the localization length is shorter, we find that the electron-photon coupling produces a stronger modification in the transmittance response, and the misfit function shows a more pronounced minimum (not shown).

To provide a more complete characterization of $\overline{ \chi }$, we consider the analysis of both $\gamma$ and $W$ simultaneously. Figure \ref{fig:contour_plot} shows a contour map of $\overline{ \chi }(\gamma,W)$ for a fixed system size $L=100$. The color scale encodes the value of the misfit function across the parameter domain $0.1<W<0.8$ and $0<\gamma<0.3$. As in the previous figures, $\overline{ \chi }$ is obtained by integrating over the energy window $0<E<2t$. 
Notice that $\overline{ \chi }$ exhibits a minimum near $(\gamma, W) \approx (\gamma_{\mathrm{true}}, W_{\mathrm{true}})=(0.15, 0.5)$, located in the intersection of the vertical and horizontal red dashed lines, demonstrating that the QIP protocol can simultaneously determine the actual values of the coupling parameter and disorder strength.

We restrict our study to the weak-to-moderate coupling regime, in which the transmission spectra still display sufficiently distinct and interpretable features for the inversion protocol to function reliably. 
In the strong-coupling regime, the pronounced redistribution of spectral weight over several photon sectors leads to overlapping structures that may prevent a controlled and unambiguous parameter reconstruction within the current transport framework \cite{CiutiCarusotto2006}.

\subsection{The Aubry-Andre-Harper model coupled to a cavity}

\begin{figure}[t]
\includegraphics[width=0.46\textwidth]{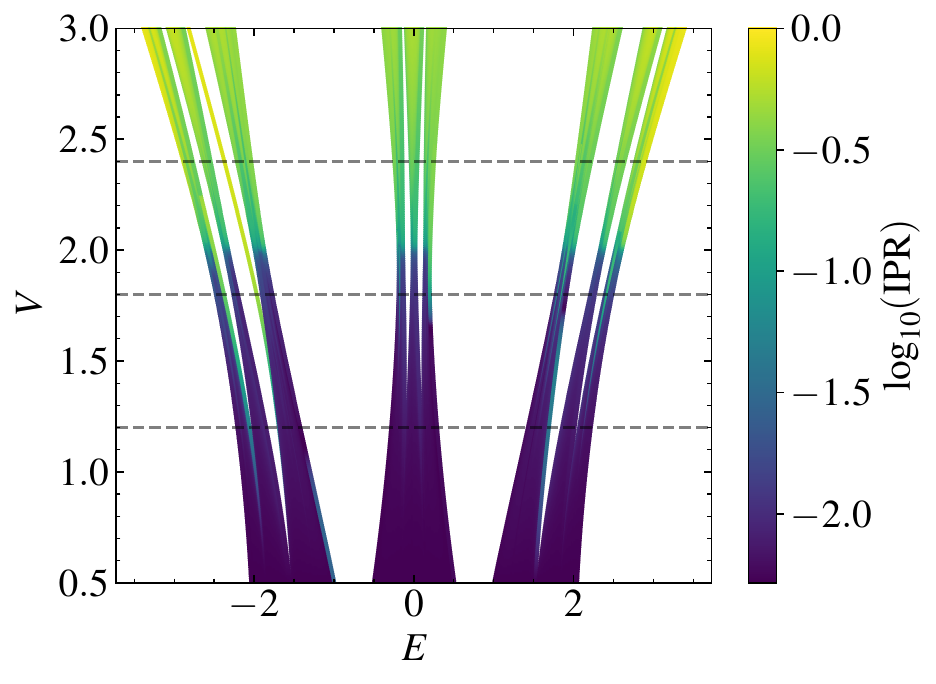}
\caption{Energy spectrum of the standard (no cavity) 1D Aubry-Andre-Harper model as a function of the quasiperiodic potential strength $V$. The color bar represents the logarithm of the inverse participation ratio, which quantifies the degree of localization. Here, we fixed $\phi = 0$, and defined $L = 300$ with open boundary conditions.}
\label{fig:aa_spectrum}
\end{figure}

Now we turn to discuss the AAH model \cite{Harper55,Aubry1980,Lahini2009}.
For the sake of completeness, we start revisiting its main properties. As mentioned in the Section \ref{AAH_section}, the AAH model exhibits a localization transition at the self-dual point $V=2t$ for all single-particle eigenstates, which is quantified by the inverse participation ratio (IPR).
In absence of a cavity and a coupling to external reservoirs, a normalized eigenstate of $\mathcal{H}_{AAH}$ can be represented as
\begin{equation}
\label{Eq:eigenstate}
\ket{\psi_{j}}=\sum_{l}\phi_{l}^{(j)}\ket{l},
\end{equation}
where $j$ labels  the $j$-th  eigenstate and $\phi_{l}^{(j)}$ is the corresponding probability amplitude in the position $\ket{l}$ basis.
Accordingly, the IPR is defined as\,\cite{Markos2006}
\begin{equation}
    \text{IPR}(\ket{\psi_{j}}) = \sum_{l}|\phi_{l}^{(j)}|^{4}~.
\end{equation}
In the thermodynamic limit ($L\rightarrow \infty$), $\text{IPR}(\ket{\psi_j})\approx L^{-1}$ $ \to 0$ for extended states, but has a finite value for localized ones.

Figure \ref{fig:aa_spectrum} shows the single-particle spectrum of the AAH model, Eq.~\eqref{eq:AAH-Hamiltonian}, as a function of the quasiperiodic potential strength $V$, for a chain of $L=300$ sites under open boundary conditions. The color bar indicates the intensity of the IPR in logarithm scale, $\log_{10}$(IPR). The figure indicates that for $V>2t$, \textit{all} eigenstates exhibit IPR $\sim 1$, characteristic of strong spatial localization, while for $V<2t$ \textit{all} eigenstates are predominantly extended, with IPR $\to 0$. The metal-insulator transition occurs at $V=2t$, with the eigenstates exhibiting multifractal properties (not shown).
The presence of a critical transition even in 1D arises from the correlated disorder potential of the model. Also, unlike the Anderson model, where the density of states forms a single disorder-broadened band, the AAH model displays multiple sub-bands separated by gaps, reflecting its quasiperiodicity. 
These features make the AAH model a useful platform to examine extended and critical states, as well as multi-band settings for the QIP protocol.

\begin{figure}[t]
\includegraphics[width=0.44\textwidth]{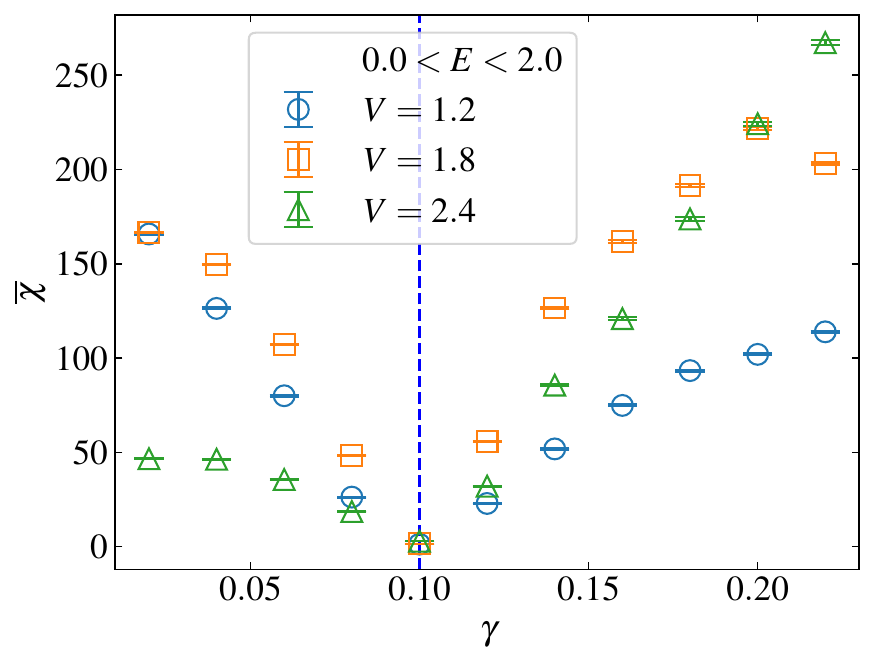}
\caption{Averaged misfit function $\overline{ \chi }$ (in arbitrary units) as a function of electron-photon coupling $\gamma$ for the AAH model, computed for three distinct quasiperiodic potential strengths $V$, and over the energy window $0 < E < 2t$. The vertical dashed line indicates the true coupling value $\gamma_{\rm true} = 0.10$.}
\label{fig:gaa_1_2}
\end{figure}

We also recall that the electron-photon coupling drastically modifies the localization properties of the eigenstates in the AAH model. The critical point at $V=2t$ broadens into a region with coexisting extended, localized, and critical states, as discussed in Ref.\,\onlinecite{macedo2024}. Therefore, in what follows, we set $\gamma_{\mathrm{true}}=0.10$ and consider three values of the quasiperiodic potential, namely, $V=1.2t$, $1.8t$, and $2.4t$. For this choice of $\gamma_{\mathrm{true}}$, these parameters place the system, respectively, in the extended, coexisting, and localized regions of the phase diagram \cite{macedo2024}.
As before, we set the energy window $\mathcal{E}_{-}=0$ and $\mathcal{E}_{+}=2t$ for the evaluation of the misfit function, with the disorder realizations being done for different values of the global phase $\phi$.

Now we turn to discuss the QIP for the AAH model coupled to a single cavity mode.
Figure \ref{fig:gaa_1_2} displays $\overline{ \chi }$ as a function of $\gamma$ for the three representative values of $V$. In all cases (extended, critical, and localized), $\overline{ \chi }$ develops a well-defined minimum at $\gamma_{\rm true}=0.1$. The most interesting feature is that the minima are roughly two orders of magnitude deeper than those for the 1D Anderson case (see, e.g., Fig.\,\ref{fig:chi_vs_gamma}). 
As this feature occurs for the three examined $V$ values, we infer that it is not related to criticality effects.
In fact, to further understand the enhanced precision of the QIP method for the AAH model, we recall an important feature of cavity QED problems: at strong light-matter coupling, $\gamma / \omega_{0} \gg 1$, the dressed quasiparticles may exhibit nonrigid-band effects\,\cite{Kockum2019}, particularly when $\hbar\omega_{0}$ is smaller than the bare electronic bandwidth, where non-perturbative approaches are required.
Therefore, multi-band systems may exhibit stronger cavity effects on conductance properties than single-band systems.

\begin{figure}[t]
\includegraphics[width=0.46\textwidth]{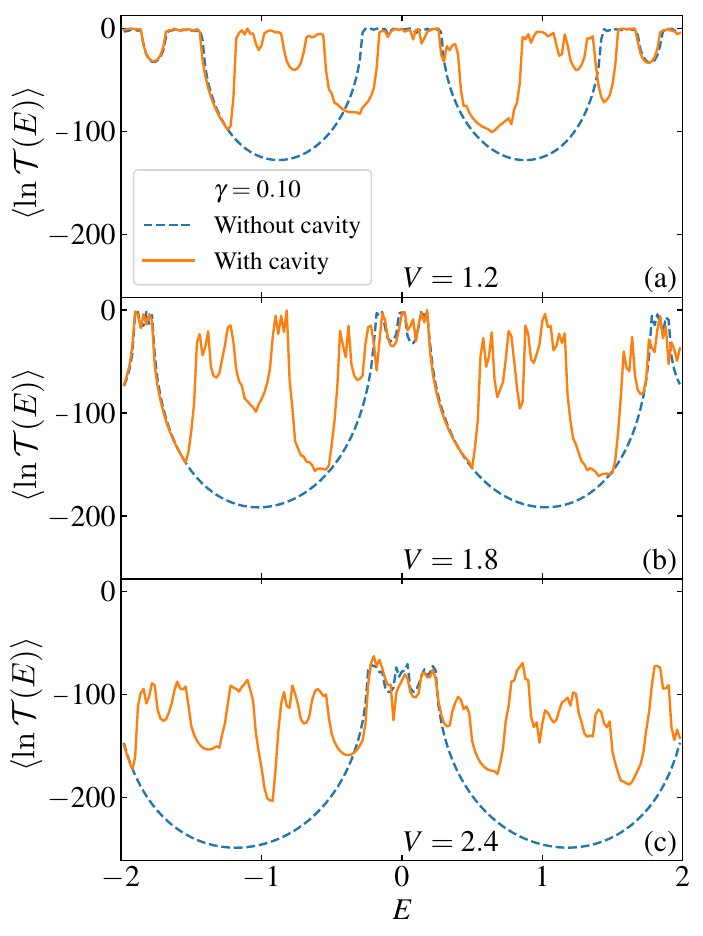}
\caption{Logarithm of the transmittance, $\ln[\mathcal{T}(E)]$, as a function of $E$ in the AAH model, for fixed $V = 1.2t$, (b) $V=1.8t$, and (c) $V=2.4t$. For each $V$, results are shown with and without coupling to the cavity mode, with fixed $\phi=0$.}
\label{fig:aa_tt}
\end{figure}

In analogy to the Anderson model results in Fig.\,\ref{fig:transmittance}, Fig.\,\ref{fig:aa_tt} details the transmittance spectra for the AAH model. 
This enables a direct comparison of the cavity-induced spectral modifications between the two disordered systems, both with and without light-matter coupling.

Figure \ref{fig:aa_tt} shows $\ln \mathcal{T}(E)$ as a function of energy $E$ for (a) $V=1.2t$, (b) $1.8t$, and (c) $2.4t$, and fixed $\phi =0$, with and without cavity effects.
In the absence of photons, the multiple band structure shown in Fig.\,\ref{fig:aa_spectrum} leads to a suppression of the transmittance within the energy intervals associated with the band gaps, as displayed by the blue dashed lines in Fig.\,\ref{fig:aa_tt}. The three fixed values of $V$ used for the transport curves are indicated by horizontal dashed lines in Fig.\,\ref{fig:aa_spectrum}, from which one may notice that the opening of a gap occurs simultaneously with a strong suppression in the transmittance response.
In stark contrast, for $\gamma=0.1$, absorption and emission photons enable photon-assisted hopping that modifies the electronic dispersion. As a result, finite transmission appears within the gaps of the bare electronic spectrum, as indicated by the orange solid curves in Fig.\,\ref{fig:aa_tt}.
Even in the localized regime, at $V=2.4t$, the coupling with the cavity modes leads to drastic changes in $\ln \mathcal{T}(E)$.
Such a strong change in transmittance is a direct and clear effect of the non-rigidity of bands in cavity QED, which, in turn, explains the enhanced sensitivity of the misfit function.

Finally, two remarks are in order. 
First, Fig.\,\ref{fig:aa_tt} supports our choice for the energy-integration bounds $\mathcal{E}_{\pm}$. Notice that, for energies near $E \approx -2t$, the transmittance with and without the cavity differs only weakly -- changes barely visible even on a logarithmic scale. At higher energies, however, $\ln \mathcal{T}(E)$ exhibits larger differences between the two cases, which increases the sensitivity of the misfit function.
Second, we comment on the transmittance in the coexistence (mixed) region, shown in Fig.\,\ref{fig:aa_tt}\,(b). Because this region hosts extended, localized, and critical states, the transmittance exhibits larger variations than in the purely extended or purely localized regions (when compared with their respective cases without cavity). Such stronger fluctuations of the transmittance increase the accuracy of the misfit function, resulting in a more distinctive minimum of $\overline{ \chi }$ when $V=1.8t$ in Fig.\,\ref{fig:gaa_1_2}, when compared to $V=1.2t$ and $2.4t$.

These findings demonstrate that systems with correlated or quasiperiodic potentials are significantly more sensitive to cavity-induced perturbations than purely random systems, making them ideal candidates for inverse-problem protocols.

\section{Discussion}
\label{Sec:discussion}

The results presented above provide a proof of concept for applying the QIP approach to conductance measurements in systems coupled to optical cavities. Although our analysis focuses on 1D systems, the procedure is general and can be extended to disordered systems in higher dimensions, since the central ingredient is the minimal coupling implemented by the Peierls substitution in the Hamiltonian. Therefore, from an experimental point of view, the theoretical models discussed in this work reach a broad class of hybrid systems where the interaction between electrons and confined electromagnetic fields can be probed via transport measurements. As cavity quantum materials (or cavitronics) are still in their infancy, in the following we discuss  current limitations and realistic perspectives for applying the QIP in experiments.

In typical circuit-QED implementations based on semiconducting devices, the cavity operates in the microwave regime, with resonance frequencies between $5-8\;\mathrm{GHz}$ and quality factors $Q$ ranging from $10^2$ to $3\times10^3$~\cite{Delbecq2011,Deng2015,Deng2016}. For instance, coplanar niobium or titanium-nitride resonators exhibit $f_c = 7.7828~\mathrm{GHz}$ and $Q \approx 3050$~\cite{Physical2018}, while carbon-nanotube devices operate near $5\;\mathrm{GHz}$ with $Q\approx 160$~\cite{Delbecq2011}. These values correspond to photon energies ($\hbar \omega_c$) of a few tens of $\mu$eV, comparable to the detuning scales observed in nanowires and quantum dots.
Despite lying in the relatively low microwave range, the cavity frequency is still high enough to couple resonantly with the electronic transitions of the quantum system and remains directly accessible to standard microwave detection techniques.

Although suitable for quantum-dot problems, an energy scale at the $\mu\mathrm{eV}$ level is very small for condensed-matter systems, where the hopping integral typically ranges from $\mathrm{meV}$ to $\mathrm{eV}$. Consequently, the ratio $\hbar \omega_{0}/t$ for the above mentioned experiments may be very small when dealing with quantum materials. Such a small value for $\hbar \omega_{0}/t$ would require nonperturbative treatments if strong or ultrastrong regimes were achieved. 
In addition, low photon energies require temperatures on the order of $\mathrm{mK}$, so that thermal effects do not mask electron-photon correlations. These considerations also constrain feasible cavity dimensions and mode volumes. For typical circuit-QED implementations, a half-wavelength cavity operating at $f_c = 6$–$8~\mathrm{GHz}$ corresponds to an effective length of a few centimeters and a mode volume on the order of $10^{-7}$–$10^{-6}~\mathrm{m}^3$. These dimensions are several orders of magnitude larger than the electronic system itself (typically a few micrometers), ensuring that the electromagnetic field is spatially uniform across the sample. This validates the long-wavelength approximation adopted in our model and confirms that the cavity behaves as a single delocalized mode interacting collectively with the 1D conductor. Such scales are consistent with the parameters reported for superconducting coplanar resonators made of niobium or titanium nitride, where quality factors $Q \sim 10^3$ and resonance frequencies near $7$–$8~\mathrm{GHz}$ are routinely achieved\,\cite{Physical2018}.

We also recall that the photon field has a spectral width characterized by the loss rate $\kappa$. Recent experiments with high-impedance TiN strips extracted $\kappa/(2\pi) \approx 2.2~\mathrm{MHz}$ for a resonator at $4.993~\mathrm{GHz}$~\cite{Wu2024}, whereas photodiode devices based on double quantum dots show cavity widths of $\kappa/(2\pi) = 15.5~\mathrm{MHz}$~\cite{Deng2015,Deng2016}. Indeed, these values correspond to quality factors from $10^2$ to $10^4$ and determine the photon lifetime in the cavity.
The finite loss rate determines the photon lifetime $\tau_{\mathrm{ph}} = 1/\kappa= Q/\omega_c$, which for a resonator with $Q = 10^4$ at $5~\mathrm{GHz}$ corresponds to $\tau_{\mathrm{ph}} \approx 300~\mathrm{ns}$. This photon lifetime is roughly $10^{2}$-$10^{4}$ times longer than the Drude scattering time of conduction electrons in metals, which typically lies between $10^{-14}-10^{-15}~\mathrm{s}$. Thus, cavities with quality factors $Q \sim 10^{2}$-$10^{4}$ can serve as viable platforms for gate-induced studies in quantum materials.
Indeed, for the implementation of the QIP protocol, the system may be modeled as approximately closed, in line with our analysis.
However, when the photon loss rate $\kappa$ is no longer negligible, theoretical descriptions based solely on a closed system become inadequate, instead requiring a treatment through the master equation to account for dissipation and decoherence\,\cite{Hagenmuller2018}. Solving such equations is often challenging, as they involve a non-unitary dynamics of the density matrix, making numerical approaches computationally demanding, and is beyond the scope of this work.

Within the context of quantum dots, the charge–cavity coupling $\gamma$ typically lies in the tens to hundreds of megahertz range: nanotube-based devices extract couplings of about $140~\mathrm{MHz}$~\cite{Delbecq2011}; in a strongly coupled silicon double quantum dot, vacuum Rabi splitting measurements yield $\gamma/(2\pi) = 175~\mathrm{MHz}$~\cite{Wu2024}; in niobium/SiGe resonators, couplings vary from $\gamma/(2\pi) \approx 21~\mathrm{MHz}$~\cite{Deng2015,Deng2016} up to 30~MHz~\cite{Petersson2012}.
That is, these experiments can access $\gamma/\omega_{0} \approx 0.1$, which is sufficient to probe the changes discussed in this work. Furthermore, these scales provide realistic values for phase shifts $\delta\phi$, corresponding to the cavity-field phase change induced by the interaction with the electronic system. Heterodyne readout techniques can detect variations smaller than $1~\mathrm{mrad}$, enabling the detection of subtle variations induced by the coupling~\cite{Delbecq2011}. The main challenge at present is to increase the number of quantum dots, although the QIP protocol appears feasible if larger systems become available. In this context, the disorder is inherent to the qubit itself.

It is also important to mention that, in experiments, the intracavity photon number may range from $n \approx 2$ photons, achieved with probe powers of $-130~\mathrm{dBm}$~\cite{Wu2024}, up to $n \sim 10^4$ photons at $-60~\mathrm{dBm}$~\cite{Delbecq2011}. Therefore, imposing a cutoff on the photon number in the theoretical treatment is acceptable in certain regimes. Furthermore, currents measured in nanowire devices under microwave irradiation lie in the picoampere to nanoampere regime; for instance, a double-quantum-dot photodiode generates a photocurrent of $2.4~\mathrm{pA}$ for $1~\mathrm{fW}$ incident power~\cite{Deng2015,Deng2016}, while cavity-coupled double quantum dots exhibit current oscillations of several nanoamperes~\cite{Physical2018}. These magnitudes define the sensitivity required of electronic amplifiers.

These energy and current scales suggest that implementing the inverse-problem protocol in cavity quantum materials is indeed experimentally feasible. In such systems, the cavity acts as a coherent, delocalized mediator whose parameters leave measurable fingerprints in the conductance. Consequently, the QIP protocol may serve as a \textit{spectroscopy} tool, enabling the extraction of key cavity parameters -- such as $\gamma$, $\kappa$, detuning $\Delta$, and finesse -- directly from transport data $\mathcal{T}(E;\gamma,\kappa,\Delta)$. 

\section{Conclusions}\label{conclusion}

In this work, we implement a nonperturbative approach to estimate electron-photon coupling in disordered systems embedded in optical cavities. As a proof of principle, we analyze one-dimensional systems, namely, the Anderson and Aubry-Andre-Harper models. In 1D, the former hosts only localized states, whereas the latter is a multi-band system that exhibits a transition between extended and localized states, with multifractal states at the transition. Therefore, this choice allows us to examine localized, extended, and critical states, as well as single and multi-band systems.

To this end, we employ an inverse problem approach, which relies on minimizing a misfit function $\chi$ defined over the transmittance spectra. 
The transmittance is computed within the nonequilibrium Green's functions formalism, which explicitly includes electron-photon coupling. 
In this way, we may extract the cavity coupling and disorder strengths of the given actual system by minimizing $\chi$ against simulated transport data, i.e.~by inverting the transmittance problem. For the two models considered, we find that the procedure successfully recovers the input parameters (disorder strength and light-matter coupling) within good numerical accuracy.

In light-matter problems, photon absorption and emission enable photon-assisted hopping that may modify the electronic dispersion. Interestingly, within the AAH model, for sufficiently large electron-photon coupling, we find a strong enhancement of the transmittance inside the bare single-particle gaps of the model. This in-gap channel modifies the overall transmittance and improves the sensitivity of the misfit function used to infer the cavity coupling. We emphasize that this mechanism does not rely on model-specific details, but is a general feature expected to occur in multi-band systems. This means that systems with several bands coupled to cavity photon modes are therefore well suited for applying the misfit function approach.

As discussed in Section \ref{Sec:discussion}, current circuit-QED and nanowire platforms can access the frequency, linewidth, and coupling ranges needed to implement the IP protocol, making a transport-based extraction of cavity parameters experimentally realistic. At the same time, scaling beyond few-dot devices, maintaining millikelvin operation with low photon loss, and controlling device-level disorder remain nontrivial and may limit parameter regimes and signal-to-noise. Thus, while near-term demonstrations appear within reach, systematic studies across larger systems will require advances in device fabrication and measurement stability.
Despite this, here we demonstrated a direct bridge between mesoscopic transport and cavity quantum electrodynamics, showing that inverse methods can retrieve optical properties from electronic observables alone, playing the role of an alternative spectroscopic tool for optical cavities.

\begin{acknowledgments}
The authors acknowledge financial support from the Brazilian funding agencies Conselho Nacional de Desenvolvimento Cient\'\i fico e Tecnol\'ogico (CNPq), Coordena\c c\~ao de Aperfei\c coamento de Pessoal de Ensino Superior (CAPES), and Fundação de Amparo \`a Pesquisa do Estado do Rio de Janeiro (FAPERJ).
N.C.C.~acknowledges support from FAPERJ Grants No.~E-26/200.258/2023 [SEI-260003/000623/2023] and E-26/210.592/2025 [SEI-260003/004500/2025], CNPq Grants No.~313065/2021-7 and 308130/2025-1, Serrapilheira Institute Grant No.~R-2502-52037, and Alexander von Humboldt Foundation.
T.F.M.~acknowledges support from FAPERJ Grant No.~E-26/202.518/2024 - SEI-260003/007717/2024.
J.F.~thanks FAPERJ, Grant No.~SEI-260003/019642/2022, and ANID Fondecyt grant No.\,3240320.
This research was partially supported by the supercomputing infrastructure of the NLHPC (CCSS210001). 
This publication has emanated from research supported in part by a research grant from Science Foundation Ireland (SFI) under Grant No. SFI/12/RC2778\_P2.
\end{acknowledgments}
\subsection{DATA AVAILABILITY}

The data and codes that support the findings of this article are openly available \cite{data}.

\color{black}

\bibliography{references}

\end{document}